\documentclass[prd,nofootinbib,amsmath]{revtex4}
 \usepackage[dvips,final]{graphicx}
  \usepackage{amssymb}
   \usepackage{amsmath}
    \usepackage{epsfig}
     \usepackage{bm}% bold math
      \usepackage{pifont}
\def\<{\langle}
\def\>{\rangle}

\begin{document}
\thispagestyle{empty}
\date{\today}
%\vspace*{-10mm}

\title{Chaotic dynamics in quark-gluon cascade}
\author{A.T. Temiraliev}
\email{abzal@sci.kz}
\affiliation{Institute of Physics and Technology \\ Almaty, Kazakhstan}
\vspace {10mm}
%\cleardoublepage
\begin{abstract}
A map to the quark-gluon cascade on the basis of nonlinearity
in the quark and gluon distributions in hadrons is proposed.
Calculations of the quarks trajectories have shown the presence of the chaotic dynamics
as a consequence of bifurcations.
\end{abstract}

\maketitle

At high energy, there are two types of the QCD processes: perturbative and non-perturbative (short and long distances, respectively). In the non-perturbative region, direct expansion in the strong coupling constant is inapplicable. Then, to describe the strong interactions, it is necessary to use in addition some assumptions and phenomenological constructions. Modern research in hadron and nuclear physics have deal more and more with nonlinear aspects of the quark-gluon dynamics. Analysis of the role of nonlinear effects in formation of the initial quark-gluon configurations in the hadrons at early stage of the nuclear impacts was given for the first time in the articles \cite{Kovch}. Later on, nonlinear effects were considered within the quasi-classical approach \cite{Leon}.

Ambiguity of the account of nonlinear effects has led to significant variety of the cascade models. Various models of the behavior of quarks and gluons have been confronted with a large amount of experimental data on the deep-inelastic scattering (DIS) structure functions (SF), which describe the quarks and gluons distribution \cite{Adl}. The dependence of quarks and gluons SF on Bjorken's variables $Q^2$ and $x$ is usually obtained from the numerical solution of the linear Dokshitzer-Gribov-Lipatov-Altarelli-Parisi (DGLAP) \cite{G-L} or Balitsky-Fadin-Kuraev-Lipatov (BFKL) equations \cite{KLF}. Evolution of the transverse-momentum dependent parton densities had been studied in \cite{Cher}.

At present, one of the greatest achievements can consider the discovery of chaotic dynamics \cite{Fej} in different areas of science. Chaotic behavior has been observed in a variety of systems including lasers, electrical circuits, fluid dynamics as well as computer models of chaotic processes. In frameworks of fractal analysis the hypothesis of universality at formation of hadrons is discussed in \cite{Bat}. After those works, it becomes clear that the dynamics at the chaos border often manifests scaling regularities. Scenario of transition to chaos means a sequence of bifurcations observed under slow variation of a control parameter on a way from regular to chaotic behavior, for example, via period-doubling cascade, quasi-periodicity, intermittency.

The evolution of dynamic system usually is described by the differential equation. However in the modern researches there is a map method allowing to describe many phenomena of nonlinear dynamics. There are a merges and splitting in the quark and gluon evolution. The collective interactions of partons (quarks and gluons) led to universality of the hadron SF. We use the renormalization-group approach to the evolution of hadron's SF, allowing to recreate a physical picture of the critical behavior. For the quark-gluon cascade, we enter an iterative map in which a number of the quarks and gluons in $(n+1)$-th generation are proportional to the number of them in $n$-th generation. To find the shares of the momentum $x$, we use the one-dimensional map for quark and gluon distributions:
\begin{equation}
x_{n+1}	= R f(x_n) \ ,
\end{equation}
where $n=0,1,2,\ldots $ and $f(x)=xq(x)$ is the momentum distribution of partons with the momentum fraction $x$ and density $q(x)$. $R$ is the parameter, determining the character of observing regimes.

The function $f(x)$ in the map $x_{n+1}=f(x_{n})$  determines evolution for one step of the discrete time. For two steps we have $x_{n+2}=f(f(x_n))$. Let us introduce instead of $x$ a new variable re-scaled with the factor $\alpha$. Changing $x$ to $x/\alpha$ in both sides of the equation, we can write $x_{n+2}=f_1(x_n)$, where $f_1(x)=\alpha f(f(x/\alpha))$. We can take $f_1(x)$ as a new initial function and perform the same operations. Then, we obtain a renormalized evolution operator for four steps: $x_{n+4}=f_2(x_n)$, where $f_2(x)=\alpha f_1(f_1(x/\alpha))$. Repetition of the procedure yields a recurrent functional equation $f_{k+1}(x)=\alpha f_k(f_k(x/\alpha))$. If the original map $f_0(x)$ depends on the parameter and demonstrates period-doubling cascade, then, at the accumulation point of the period-doubling. The limit function f(x) will satisfy to the equation $f(x)=\alpha f(f(x/\alpha))$ \cite{Tur}. It is a fixed point of the functional equation.

The momentum distributions of quarks in a hadron for the momentum fraction $x \rightarrow 1$ possess, according to the QCD sums rules, asymptotic behavior of the valence quarks which define the quantum numbers of hadron:

$$ 
\lim f(x)\rightarrow (1-x)^{2n_v-1} \ ,
$$
where $n_v$ is the number of the valence quarks between which the rest of the hadron momentum ($n_v=2$ for nucleons and $n_v=1$ for mesons) is distributed. It is yet not possible to prove asymptotic formulas in limits of QCD for partons with small fraction of hadron momentum $(x\ll 1)$ at Regge region. We use simple parameterization of the SF from  experimental data at great values of 4-momentum transferred at interaction ($Q^2\geq  10 \ Gev^2$) for the proton:
$xu_v(x)=1.8x(1-x)^{2.5}$, $xd_v(x)=3.6x(1-x)^{1.5}$, $xS(x)=0.1(1-x)^6$, $xG(x)=2.76(1-x)^5$, where $u_v$, $d_v$ are the valence quark, $S$ is the sea quark, and $G$---the gluon distributions. The momentum distribution of the quarks reads:
\begin{equation}
f(x) 
= 
1.8x(1-x)^{2.5}+3.6x(1-x)^{1.5}+0.1(1-x)^6 \ . 			
\end{equation}

Thus, the quark momentum fraction in the nucleon is 0.54 and the gluon one is almost one half 0.46. Positive terms in the equation (1) with $f(x)$ from (2) describe the increase of the partons number and negative their is the reduction, i.e.  quark-antiquark,  quark-gluon and  gluon-gluon recombination.

Numerical solution of the equation (1)  has shown that there is the evolution termination in the field of small values of evolution parameter  ($R < 0.3$) at any initial value $x_0$ of parton's momentum fractions. At increasing of the  parameter $R$ the transition is in established  mode that corresponds to long enough cascade process at great values of iterations. It is observed that the number of partons do not change. Trajectories ($x_n$)  after transient reach some steady value (motionless point). At $R=0.8$ after transient the display orbit becomes strictly periodic with the period two (bifurcation of motionless point). At $R> 0.9$ the orbit represents more difficult movement. These processes are shown in Figure 1
%\ref{}3regime}
 at $R=0.2$, $R=0.8$ and $R=0.99$. At the fixed value $R$ and changing initial values of momentum fractions  $x_0$ of the partons  after  transition period  don't depend on the initial value $x_0$.

\begin{figure}[htp]
\begin{center}
	%\centering
				\resizebox{\textwidth}{!}{
			\includegraphics[totalheight=1in]{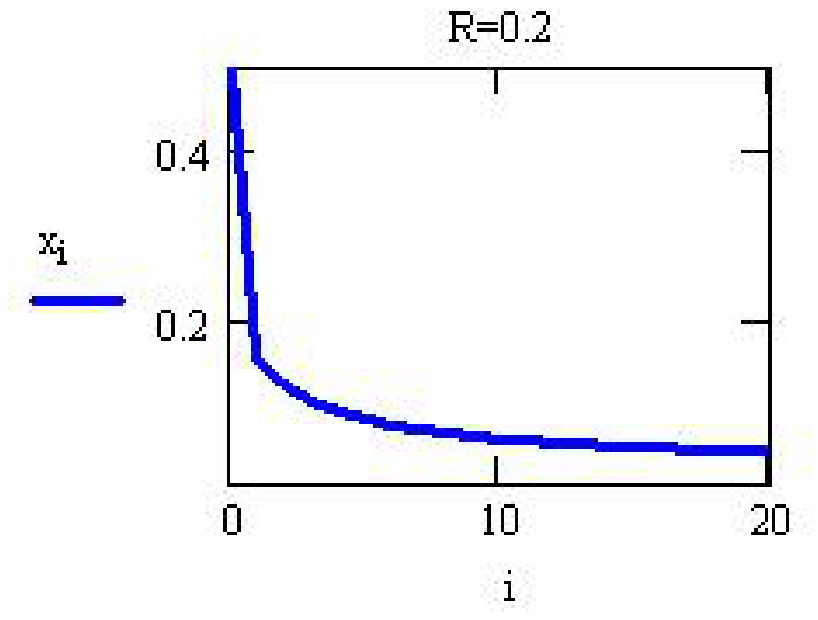}
		\includegraphics[totalheight=1in]{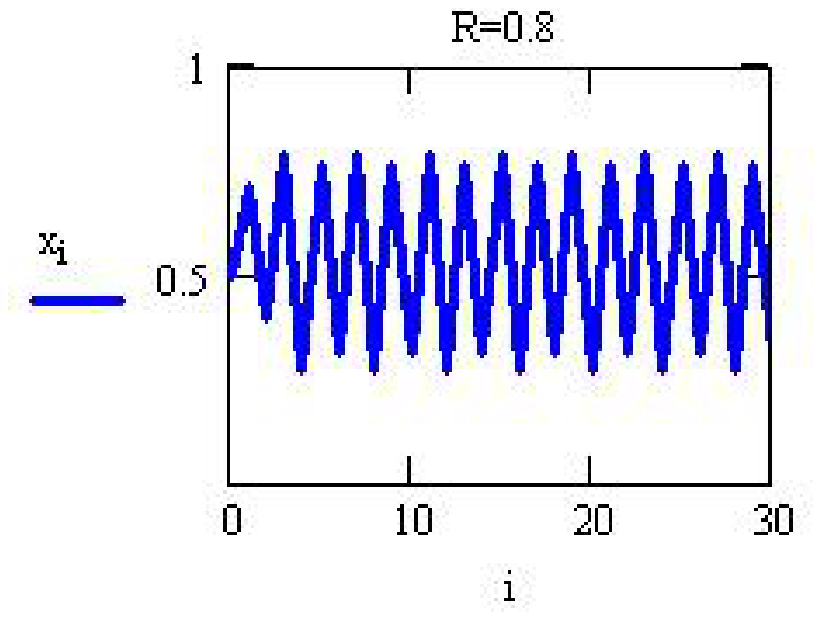}
				\includegraphics[totalheight=1in]{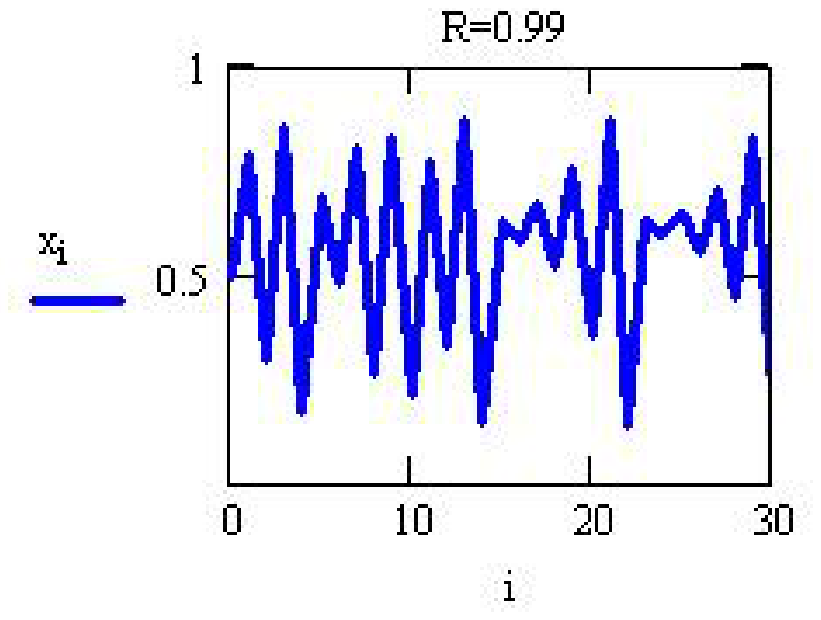}
				}
	\caption{R=0.2 R=0.8 R=0.99}
%\label{3regime}
\end{center}
\end{figure}

The change of trajectory character most visually shows the dependence schedule $x_{i,j}=f(x_{i-1,j},R_j)$ at iterations of parameter of evolution $R_j=0.99j/500=0\div 0.99$, where $i$ is the number of points on variable $x$ (number of iterations), $j$ is the number of values of the variable $R_j$. At large values of $R$, the system transfers in a chaotic mode when two close points run up on different trajectories that is shown in Figure 2 is observed. Thus, casual small initial changes of the system of the cooperating partons can lead to as much as big changes of the dynamics of the system.

\begin{figure}[htp]
\centering
	\resizebox{\textwidth}{!}{
		\includegraphics[width=\textwidth]{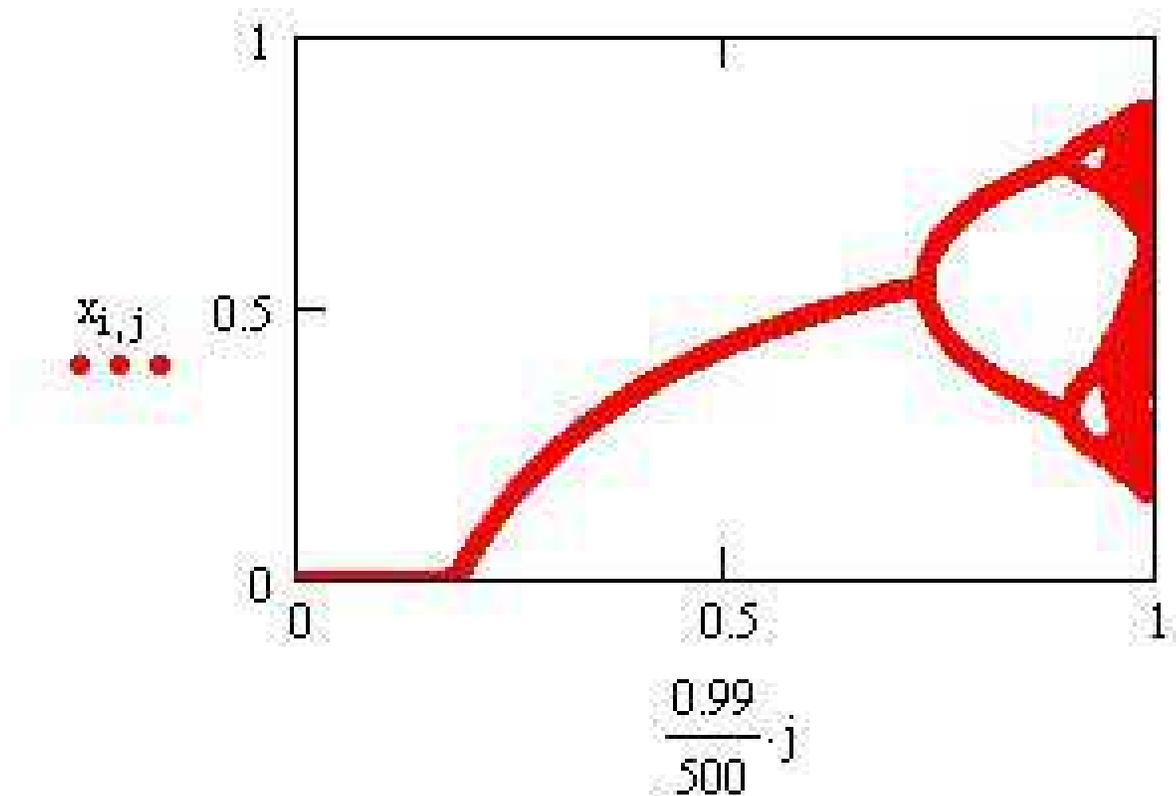}
				}
			\caption{Attracts, chaotic dynamics}
%	\label{figure2}
\end{figure}

 In order to distinguish between the chaotic and not chaotic modes, we compare orbits to close entry conditions in these modes. As a measure of this difference, we choose the difference module between values of corresponding orbits of the display, carried to the value of one of the orbits.

\begin{figure}[htp]
	%\centering
	\resizebox{\textwidth}{!}{
		\includegraphics[totalheight=1in]{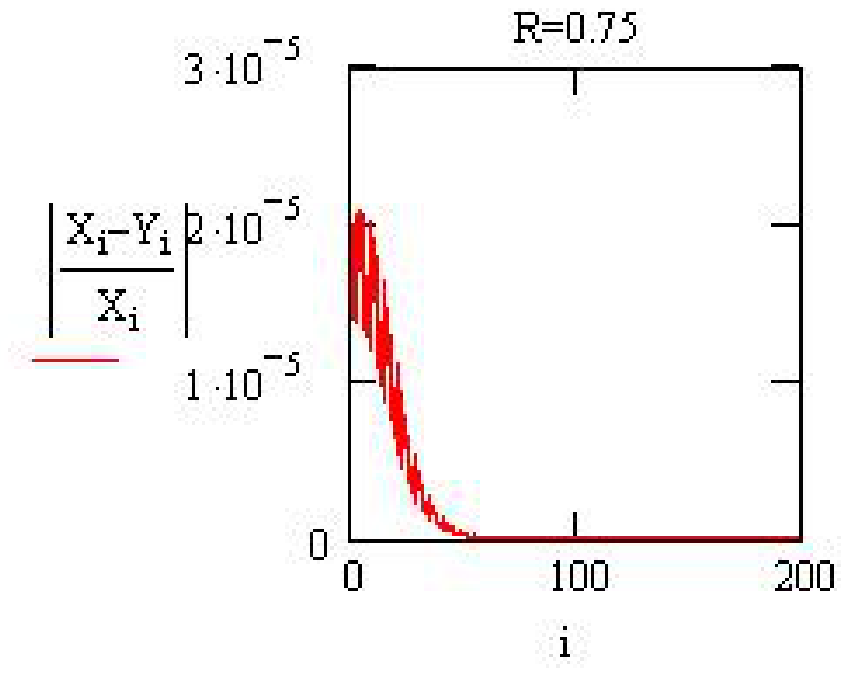}
		\includegraphics[totalheight=1in]{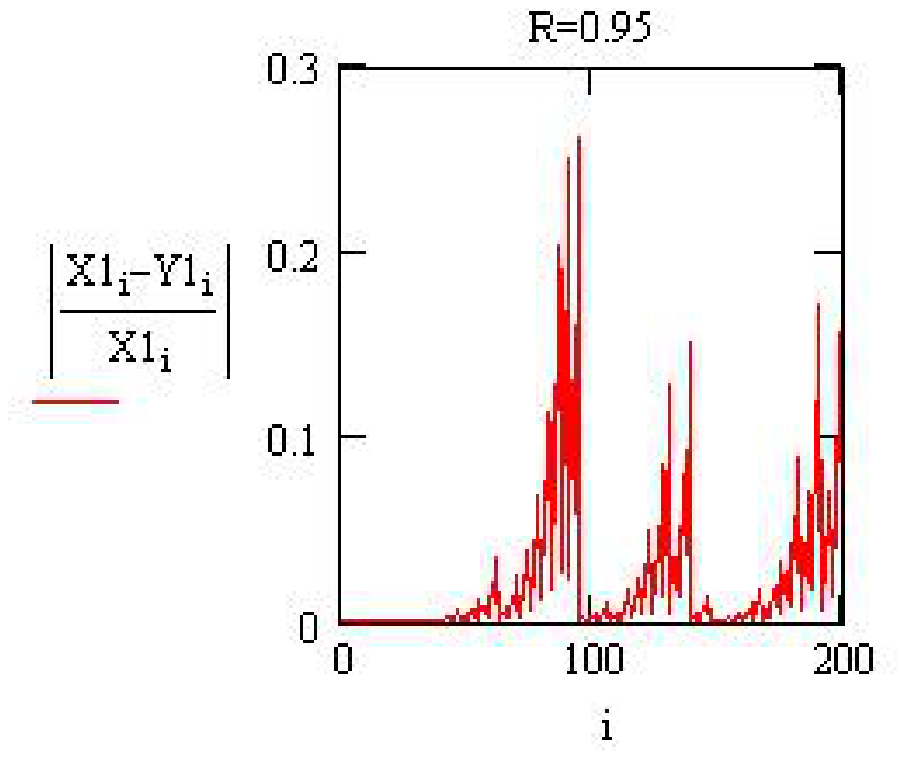}
		}		
			\caption{Regular and chaotic modes}%	\label{figure3}
\end{figure}

At initial approach for the first trajectory $x_0 =0.5$ and $y_0=0.50001$ for the second one. Results of calculations of trajectories for $R=0.95$ and $R=0.75$ dynamics are presented in Figure 3.
%In \cite[\$5]{Tur} had shown

\textbf{Conclusion:} In the quark-gluon cascade, the essential role is played by not only linear parts of the parton density, but also the higher order terms. Dynamic quark-gluon systems are highly sensitive to the initial conditions. The strange quark trajectories (attracts) display a new nonlinear phenomenon in the hadron physics --- dynamic chaos in the quark-gluon evolution. Scenario of the transition to the chaos means a sequence of bifurcations observed under the slow variation of the control parameter.

\textbf{Acknowledgments:} I am grateful to I.O. Cherednikov for his attention to the article and help.

\bibliographystyle{alpha}

\end{document}